\documentclass[a4paper,12pt]{article}
\usepackage{amssymb,amsmath,amscd,epsfig,bbold,stmaryrd,bm}
\title{A necessary condition for the thermalization of a quantum system coupled to a quantum bath}

\author{
 Oleg Lychkovskiy
\thanks{e-mail: lychkovskiy@itep.ru}
\\
{\small\it Institute for Theoretical and Experimental Physics}\\
{\small\it 117218, B.Cheremushkinskaya 25,
Moscow, Russia}
}

\date{}

\begin{document}
\maketitle

\newcommand{\be}{\begin{equation}}
\newcommand{\ee}{\end{equation}}

\newcommand{\ii}{\mathrm{i}}
\newcommand{\pp}{\mathbf{p}}
\newcommand{\ssigma}{\bm{\sigma}}
\newcommand{\q}{\mathbf{q}}
\newcommand{\e}{\mathbf{e}}
\newcommand{\A}{\mathcal{A}}

\newcommand{\p}{\tilde{p}}
\newcommand{\x}{\tilde{x}}

\newcommand{\la}{\langle}
\newcommand{\ra}{\rangle}

\newcommand{\cH}{{\cal H}}
\newcommand{\cS}{{\cal S}}
\newcommand{\cB}{{\cal B}}
\newcommand{\cF}{{\cal F}}

\newcommand{\hPS}{ P^{\cal S}}
\newcommand{\hPE}{ P^{\cal E}}
\newcommand{\hH}{ H}
\newcommand{\hHB}{ H^{\cal B}}
\newcommand{\hHS}{ H^{\cal S}}
\newcommand{\hHSB}{ H^{\cal SB}}
\newcommand{\hV}{ V}
\newcommand{\hS}{ S}
\newcommand{\psiS}{\psi^{\cal S}}
\newcommand{\psiB}{\psi^{\cal B}}

\newcommand{\dens}{ \rho}
\newcommand{\densS}{ \rho^{\cal S}}
\newcommand{\densB}{ \rho^{\cal B}}
\newcommand{\meandens}{{ \bar \rho}}
\newcommand{\meandensS}{\overline{{ \rho}^{\cal S}}}
\newcommand{\meanavdensS}{\langle \overline{{ \rho}^{\cal S}}\rangle_{\cal B}}

\newcommand{\tr}{\mathrm{tr}}
\newcommand{\trS}{\mathrm{tr}_{\cal S}}
\newcommand{\trE}{\mathrm{tr}_{\cal E}}
\newcommand{\Pu}{\mathfrak P}

\newcommand{\im}{\mathrm{i}}

\begin{abstract}
A system put in contact with a large heat bath normally thermalizes. This means that the state of the system $\densS(t)$ approaches an equilibrium state $\densS_{\rm{eq}},$ the latter depending only on macroscopic characteristics of the bath (e.g. temperature), but not on the initial state of the system. The  above statement is the cornerstone of the equilibrium statistical mechanics; its validity and its domain of applicability are central questions in the studies of the foundations of statistical mechanics. In the present paper we concentrate on one aspect of thermalization, namely, on the system initial state independence (ISI) of $\densS_{\rm{eq}}.$  A necessary condition for the system ISI is derived in the quantum framework. We use the derived condition to prove the {\it absence} of the system ISI in a specific class of models. Namely, we consider a single spin coupled to a large bath, the interaction term commuting with the bath self-Hamiltonian (but not with the system self-Hamiltonian). Although the model under consideration is nontrivial enough to exhibit the decoherence and the approach to equilibrium, the derived necessary condition is not fulfilled and thus $\densS_{\rm{eq}}$ depends on the initial state of the spin.
\end{abstract}

\section{Introduction}

In the last decade a considerable progress in deriving the fundamentals of statistical physics from the first principles of quantum theory was achieved 
\cite{Tasaki1998}-\cite{Gogolin} (see \cite{Gogolin} for a profound list of references to the related works).
Most of the results which contribute to this progress were  obtained in the following framework. Consider a quantum system (described by a Hilbert space $\cS$), which interacts with a quantum bath (described by a Hilbert space $\cB$). The bath is considered to be "much larger" than the system. In particular, if $\cS$ and $\cB$ are finite-dimensional with the dimensions $d_{\cS}$ and $d_{\cB}$ correspondingly, which is assumed in what follows, then $d_{\cS}\ll d_{\cB}.$  The composite system with Hilbert space $\cH=\cS \otimes \cB$ is considered to be closed and to evolve according to the Shroedinger equation with a Hamiltonian
\be
\hH=\hHS+\hHB+\hHSB,
\ee
where $\hHS$ and $\hHB$ are self-Hamiltonians of the system and the bath correspondingly, and $\hHSB$ is an interaction Hamiltonian.
Here and in what follows the usage of superscripts and subscripts $\cS,~\cB$ is believed to be self-explanatory.
%
%
The state of the combined system $\cH$ is described by a state vector $\Psi\in\cH.$ The latter evolves as $\Psi(t)=\exp(-\ii \hH t)\Psi(0).$  The states of the system $\cS$ and the environment $\cB$ are described by the reduced density matrices
\be\label{densS}
\densS \equiv \tr_{\cB} |\Psi\ra\la\Psi|,~~~\densB \equiv \tr_{\cS} |\Psi\ra\la\Psi|
\ee
correspondingly.

The initial state of the composite system $\cH$ is taken to be a product state:
\be\label{initial state}
\Psi(0)=\psi\Phi,~~~\psi\in\cS,~~~\Phi\in\cB.
\ee
Here and in what follows we use Greek letters $\Psi,~\psi$ and $\Phi$ to denote the normalized vectors of unitary spaces $\cH,~\cS$ and $\cB$ correspondingly. The product form of the initial state is natural when considering the approach to the thermal equilibrium. Usually all results obtained with the use of the product initial state assumption may be generalized to the case of an arbitrary pure initial state. Moreover, usually the results obtained for a pure state $\Psi$ are straightforwardly generalized to the case when the state of the composite system $\cH$ is mixed and described by the density matrix $\dens.$
%
%





What can be said about the long-time behavior of $\densS(t)$ in the case of a generic interaction $\hHSB?$  This is a central questions of the equilibrium statistical mechanics. An intuitive answer is that $\densS$ approaches an equilibrium density matrix of some special (e.g. canonical) form. As  was argued in \cite{Linden2009}, on closer examination one expects that the system exhibits four distinct properties, which we refer to as {\it thermalization properties.}
We formulate them below exploiting the product form of the initial state (\ref{initial state}):
\begin{enumerate}
    \item{{\it Equilibration.} By definition, a system {\it equilibrates} if $\densS(t)$ approaches a time-averaged density matrix $\meandensS$ and stays close to it most of the time. Defined in this way equilibration does not imply neither any special form of $\meandensS,$ nor the independence of $\meandensS$ from initial conditions.}
    \item{{\it Bath initial state independence (Bath ISI).} This means that $\meandensS$ (almost) does not depend on the exact microstate of the bath, $\Phi.$ Rather $\meandensS$ should depend on some macroscopic characteristics of the state of the bath, which should be represented by functionals on $\cB.$ The prime example of such characteristic is the bath inverse temperature $\beta=\beta[\Phi].$}
    \item{{\it System initial state independence (System ISI).} This means that $\meandensS$ (almost) does not depend on $\psi.$}
    \item{{\it Boltzmann-Gibbs form of the equilibrium state:} $\meandensS=Z^{-1}\exp(-\beta \hHS).$ This property may be expected if the interaction $\hHSB$ is in some sense "weak" compared to the system self-Hamiltonian $\hHS$ (although "sufficiently strong" to ensure {\it equilibration}) and the initial state of the bath has a small energy uncertainty.}
\end{enumerate}
The last three properties make sense only if the firsts one holds. The last property makes sense if also the properties (2) and (3) hold. Note the lack of the symmetry between the definitions of the bath ISI and the system ISI. This asymmetry arises because the bath is assumed to be much larger than the system.

The ultimate goal is to derive all four properties from the first principles of quantum theory under reasonable conditions.

The present paper addresses the third property. To start with, we briefly review the main results concerning all four properties.
The first key fact was discovered in \cite{Tasaki1998}-\cite{Goldstein2006}.  It is based on the concentration of measure phenomenon, which is a striking feature of geometry in spaces of very high dimensions. Consider a linear subspace $\cH_R\subset\cH$ with dimensionality $d_R\gg 1.$ Then for almost all states  $\Psi\in\cH_R$ the reduced density matrix $\tr_{\cB} |\Psi\ra\la\Psi|$  is close to the averaged over $\cH_R$ matrix:
\be \tr_{\cB} |\Psi\ra\la\Psi| \simeq \la~ \tr_{\cB} |\Psi\ra\la\Psi| ~\ra_{\cH_R}. \ee
Usually the trajectory $\Psi(t)$ entirely lies in some $\cH_R,$ for example in the energy shell, which is spanned by the eigenvectors of $\hH$ with the eigenvalues in some range $[E,E+\Delta E].$ In this case it is natural to {\it assume} that $\densS(t)\equiv\tr_{\cB} |\Psi(t)\ra\la\Psi(t)|$ for almost every $t$ is close to the  $\cH_R$-average   $\la~ \tr_{\cB} |\Psi\ra\la\Psi| ~\ra_{\cH_R}.$
Such assumption implies that the property (1) generally holds; the properties (2) and (3)  also hold provided that all the considered initial states belong to $\cH_R.$ Also, under certain assumptions, one may perform the averaging over $\cH_R$ explicitly to get the Boltzmann-Gibbs form of the averaged reduced density matrix, which addresses the property (4).

Although the above arguments provide an important insight into the problem, they do not constitute the proofs of properties (1) -- (4). The reason is that the trajectory $\Psi(t)$ actually never completely fills any high-dimensional linear subspace $\cH_R.$ Rather it generically fills some torus \cite{von Neumann}. The dimensionality of the latter depends on the rate of the entanglement between the bath and the system introduced by the interaction $\hHSB.$  In fact, the entanglement appears to be of a primary importance in the problem involved.
In case of non-interacting system and environment ($\hHSB=0$), when the entanglement is completely absent, none of the properties (1) -- (4) hold. They do not also hold in case of non-zero but very weak interaction, when the operator norm of $\hHSB$ is much smaller than the typical energy level spacing of the environment. In the latter case the perturbation theory may be used to calculate the eigenstates and eigenvalues of $\hH$ and to demonstrate that  $\densS(t)$ does not equilibrate.

Substantial success in derivation of the properties (1), (2) and, partly, (3) was achieved in \cite{Linden2009}. A general quantum-mechanical closed system divided in two parts was considered; the only requirement for the total Hamiltonian $\hH$ was the nondegeneracy of energy gaps.   First, it was proven that the equilibration property holds provided the initial state $\Psi(0)$ is a superposition of a large number of eigenvectors of the total Hamiltonian $\hH.$ Second, it was shown that the equilibrium density matrix $\meandensS$ is almost the same for almost all initial states $\Phi$ of environment which belong to a high-dimensional subspace $\cB_R\subset\cB;$ those states of the bath which provide the exceptions from this rule form a subset in $\cB_R$ of an exponentially small measure. Thus, the bath ISI property was proven. Third, an inequality was derived which in certain circumstances (in particular, when $d_{\cS}\gg 1$) proves the system ISI.
However, as  was emphasized in  \cite{Linden2009}, the problem of the system state independence is rather complicated. In particular, it was pointed out in \cite{Gogolin} that the inequality derived in \cite{Linden2009} is not very restrictive when the dimensionality of the Hilbert space of the system $\cS$ is low.

Among other important advances one should mention the strong upper bounds on the speed of fluctuations around the equilibrium state derived in \cite{Reimann2008,Linden2009arXiv}.

In the present paper the system initial state independence problem is addressed, especially in the case when $d_{\cS}$ is small. In particular, a single spin $1/2$ is considered as a system ($d_{\cS}=2$). The paper is organized as follows. In section \ref{sect definitions} we introduce definitions and notations. In section \ref{sect sufficient condition} we quote and discuss a key theorem from \cite{Linden2009} which allows to prove the bath ISI and, in some cases, the system ISI.   In section \ref{sect necessary condition} we present our main results concerning the system ISI property. 
In section \ref{sect specific model} a class of exactly solvable models is considered for which the derived necessary condition is not satisfied and thus the system ISI property does not hold.
The outline of our results is presented in section \ref{sect summary}.




\section{\label{sect definitions}Definitions}

Let us start from introducing the setup, definitions and notations.

Any finite system returns to the arbitrarily small vicinity of its initial state infinitely many times (although the recurrence time is normally very large).  For this reason the limit
$
\lim\limits_{t \to \infty} \densS(t)
$
does not exist. Instead a time-averaged density matrix
\be\label{meandensS}
\meandensS\equiv \lim_{t \to \infty}\frac{1}{t}\int_{0}^t \densS(t')dt'.
\ee
should be considered. Note that throughout the paper we use an overline to denote the time-averaging, and angle brackets to denote the averaging over normalized vectors from some subset of a Hilbert space with a uniform measure, the latter being defined in the end of the present section.

To quantify the difference between two states $\dens_1$ and $\dens_2$ we use the  distance $||\dens_1-\dens_2||,$ where
\be
||\dens||\equiv \tr\sqrt{\dens^2}.
\ee
The maximal value of this distance is 2. This is a physically meaningful definition, as it is discussed e.g. in \cite{Linden2009,Kupsch2003}. In particular, it is equal to the doubled
maximum difference in probability for any outcome of any
measurement performed on the two states \cite{Linden2009}.
\footnote{Note that the above defined distance is the same as in \cite{Popescu2006} but is two times larger than in \cite{Linden2009}. The accepted definition is natural when considering the distance between the states of a single spin, see below.}

The total Hamiltonian is assumed to be nondegenerate,
\be
\hH=\sum_{n=1}^d E_n |\Psi_n\ra\la\Psi_n|,
\ee
where $\Psi_n$ are the eigenvectors of $\cH,$ and $E_n<E_m$ for $n<m.$ The initial sate of the composite system is a superposition of the eigenvectors with coefficients $c_n:$
\be
\Psi(0)=\sum_{n=1}^d c_n \Psi_n.
\ee

The time evolution of $\densS(t)$ reads
\be
\densS(t)=\sum_{n=1}^d \sum_{m=1}^d  c_n c_m^* e^{-\ii (E_n-E_m)t} \densS_{nm},
\ee
where the matrices
\be
\densS_{nm} \equiv \tr_{\cB}|\Psi_n\ra\la \Psi_m|
\ee
are introduced (not to be confused with matrix elements!). Evidently, these matrices encode the dynamics of the open system $\cS,$ while the coefficients $c_n$ describe the initial conditions. The nondegeneracy of the Hamiltonian allows to obtain
\be\label{meandensS expanded}
\meandensS=\sum_{n=1}^d  |c_n|^2 \densS_n,
\ee
where a short-form notation $\densS_n\equiv\densS_{nn}$ is used. The time-averaged state of the system $\meandensS$ depends, in general, on the initial states of the system and the bath, $\psi$ and $\Phi$ correspondingly, through the coefficients $c_n=\la\Psi_n|\psi\Phi\ra:$ $\meandensS=\meandensS[\psi\Phi].$

In the case when the system $\cS$ is represented by a single spin, any $\densS$ may be parameterized by a polarization vector $\pp,$
\be
\densS=(1+\pp\ssigma)/2, ~~~\pp=\tr_{\cS}(\densS\ssigma),~~~0\leq|\pp|\leq 1
\ee
The polarization vector belongs to a unit sphere which is known as the Bloch sphere. The length of a polarization vector equals 1 for a pure state and is less than 1 for a mixed state. The distance between two states $\densS_1$ and $\densS_2$ is simply the Euclidian distance in the Bloch sphere:
$$||\densS_1-\densS_2||= |\pp_1-\pp_2|\leq 2.$$
The scalar product of vectors $\pp$ and $\pp'$ is denoted as $(\pp,\pp').$ We define the following important polarization vectors: the initial polarization vector $\pp_0=\la\psi|\ssigma|\psi\ra,$ polarization vectors which correspond to eigenstates of composite system $\pp_n=\la\Psi_n|\ssigma|\Psi_n\ra$ and the time-averaged polarization vector $\overline\pp=\sum\limits_{n=1}^d  |c_n|^2 \pp_n.$

In order to introduce averages and to formulate propositions about states which are typical for some subspace $\cH_R\in\cH,$ we need to define a uniform measure on $\cH_R.$  Strictly speaking, pure states of a physical system are in one-to-one correspondence with one-dimensional linear subspaces
of a Hilbert space or, equivalently, with rank one projectors $|\Psi\ra\la\Psi|.$  Therefore actually one should consider the projective space $\cH_R P$ instead of the Hilbert space $\cH_R.$ It is possible to define a uniform measure on a projective space of pure quantum states through the Haar measure on a $SU(d_R)$ group, taking into account that any pure state $|\Psi\ra\la\Psi|$ may be obtained from some fixed state $|\Psi_0\ra\la\Psi_0|$ by the unitary transformation (see e.g. \cite{Gogolin} for the details). However, following \cite{Popescu2006,Linden2009} we use a different, more explicit construction, which leads to the same result. Namely, let us choose an arbitrary basis $\{\Psi_l\}$ in $\cH_R$ and establish a map
\be
\Psi\leftrightarrow x\in \mathbb{R}^{2d_R}:~~~ x_{2l-1}={\rm Re} \la\Psi_{l}|\Psi\ra,~ x_{2l}={\rm Im} \la\Psi_{l}|\Psi\ra.
\ee
All normalized vectors  from $\cH_R$ are therefore in one-to-one correspondence
with points of the $2d_R-1$-dimensional unit sphere embedded in the $2d_R$-dimensional Euclidian space. Note however that a physical state $|\Psi\ra\la\Psi|$ corresponds not to a point but to a one-dimensional curve on the sphere, because of the overall phase ambiguity of $\Psi.$ Now to pick up a quantum state from  $\cH_R$ (more precisely, from  $\cH_RP$) at random according to the uniform measure, we first pick up a vector $x$ from a unit sphere according to the uniform measure on a sphere, and then construct the corresponding state $|\Psi\ra\la\Psi|.$ A thereby constructed measure does not depend on the choice of the basis  $\{\Psi_l\}.$

In the above paragraph we reminded a well-known fact that actually a pure physical state should be characterized by a projector $|\Psi\ra\la\Psi|$ (or by a vector $\Psi$ ``up to a phase factor''). Bearing this in mind, in what follows we use a common language and speak about ``state vectors'' $\Psi$  and  ``state spaces'' $\cH,\cH_R,...$ without further stipulations.

\section{\label{sect sufficient condition}
Sufficient condition
for the system initial state independence}

The following theorem concerning
the initial state independence was proven in \cite{Linden2009}
:\\
{\bf Theorem 0.} Consider the hamiltonian $\hH$ with nondegenerate energy gaps, which means that $E_k-E_l=E_m-E_n$ implies either $k=l,~ m=n,$ or $k=m,~ l=n.$  \\
(i) Almost all initial states chosen from a large restricted subspace $\cH_R\subset\cH$ with the dimensionality $d_R$ yield the same equilibrium state. In particular,
\be\label{bath state independence}
\left\la\left|\left|~\meandensS-\la\meandensS\ra_{\cH_R}~\right|\right|\right\ra_{\cH_R}
\leq \sqrt{\frac{d_{\cS}\delta}{d_R}}
\leq \sqrt{\frac{d_{\cS}}{d_R}}~
\ee
with
\be
\delta \equiv \sum_{n=1}^d \la\Psi_n|(d_R)^{-1}\Pi_{\cH_R}|\Psi_n\ra ~\tr_{\cS}(\densS_n)^2 \leq 1,
\ee
where $\Pi_{\cH_R}$ is the projector onto $\cH_R.$\footnote{In general, the initial condition is not required to be of a product form in this theorem.}\\
(ii) There are exponentially few states in $\cH_R$ which yield a substantial distance between  $\meandensS$ and $\la\meandensS\ra_{\cH_R}.$ In particular, for a random state $\Psi\in\cH_R$
\be
\Pr\left\{ \left|\left|~\meandensS-\la\meandensS\ra_{\cH_R}~\right|\right|> \sqrt{\frac{d_{\cS}\delta}{d_R}}+\epsilon\right\}\leq 2 e^{-c d_R \epsilon^2},
\ee
where $c=1/(18\pi^3).$


First we review how this theorem may be used to prove the bath ISI property \cite{Linden2009}. Let us choose any system state $\psi$  and consider $\cH_R$ as a tensor product of $\psi$ and some large $\cB_R\subset \cB$ with the dimensionality $d_R\gg d_{\cS}:$  $\cH_R=\psi\otimes\cB_R.$ Then
one gets that for a fixed $\psi$ and vast majority of $\Phi\in\cB_R$ the equilibrium state $\meandensS$ is
close to the average $\la\meandensS\ra_{\cB_R}.$
In other words, it is proven that for any fixed initial state $\psi$ of the system the equilibrium state $\meandensS$ depends on the initial state of the bath $\Phi\in\cB_R$ extremely weakly. 

Note that the smallness of $\delta$ is not required in the above proof; in fact one may safely take $\delta=1$ and exploit the weaker bound in (\ref{bath state independence}).  The bound with $\delta$ was introduced in  \cite{Linden2009} in order to treat the system ISI problem. The latter appears to be more complicated compared to the previous one. Indeed, let us try to proceed analogously to what was done in the preceding paragraph. We fix some state of the bath $\Phi$ and construct $\cH_R = \cS\otimes \Phi$ (according to the formulation of the system ISI property we should take the whole $\cS$ instead of some subspace in $\cS$). Now, however, $d_R=d_{\cS},$ and the weaker bound in (\ref{bath state independence}) appears to be useless. The stronger bound is useful provided $\sqrt{\delta}$ is small. For this reason one may look at Theorem 0 with $\cH_R = \cS\otimes \Phi$ as on the\\
\newline
{\bf Sufficient conditions for the system ISI:} If
\be\sqrt{\delta}\ll 1,\ee
then the system ISI property holds.
\newline

In \cite{Linden2009} a case was considered when, firstly, the dimensionality of the system is large, $\sqrt{d_{\cS}}\gg 1,$ and, secondly, the eigenstates $\Psi_n$ are highly entangled (in particular, far from product), which implies that the purities of the density matrices $\densS_n$ are close to their minimal values:
\be\label{minimal purity}
\tr_{\cS}(\densS_n)^2\simeq 1/d_{\cS}.
\ee
In this case $\delta\simeq 1/d_{\cS},$ the above sufficient condition is satisfied and the system ISI property is thus proven.


We emphasize, however, that if the dimensionality of the system is small, the above condition can not be satisfied,\footnote
{
After the present article was completed and submitted to arXiv, we learned about a very recent previous work by Christian Gogolin \cite{Gogolin}, in which he expressed the same criticism concerning the uselessness of the result of \cite{Linden2009} in the case of small $d_{\cS}.$ Moreover, he proved another sufficient conditions for the system ISI, which works well for small $d_{\cS},$ but relies on the {\it eigenstate thermalization hypothesis.} The latter is discussed in the following section.
} as
\be
\delta \geq 1/d_{\cS}.
\ee
In particular, for a single spin $1/2$ considered as a system one {\it at best} obtains from (\ref{bath state independence})  
\be
\left\la\left|\left|~\meandensS-\la\meandensS\ra_{\cS}~\right|\right|\right\ra_{\cS}\leq 1/\sqrt{2},
\ee
which is not very restrictive.
Thus in the case of small $d_{\cS}$ Theorem 0 does not answer the question whether the system ISI property holds or not. Evidently in this case the system ISI problem requires some additional treatment. In the following section we derive a {\it necessary condition} for the system initial state independence, which in particular appears to be useful when $d_{\cS}$ is small.

\section{\label{sect necessary condition}Necessary condition for the system initial state independence}

First let us refine the definition of the system initial state independence.\\
{\bf Definition.} The equilibrium state $\meandensS$ of the system is independent from the initial state of the system $\psi$ for a fixed initial state of the bath $\Phi$ with the accuracy $\varepsilon$ if
\be\label{independence inequality}
\left|\left|~\meandensS[\psi\Phi]-\la\meandensS\ra_{\cS}[\Phi]~\right|\right|<\varepsilon~~~\mathrm{for~any}~~\psi.
\ee
We remind that $\la...\ra_{\cS}$ denotes the averaging over the normalized states from $\cS$ with a uniform measure, while brackets $[...]$ indicate the functional dependence.

According to the above definition, to prove the system ISI property means to establish the inequality (\ref{independence inequality}) for (almost) any initial bath state $\Phi$  with some small $\varepsilon$ under  reasonable conditions. In the present paper we do not provide such a proof. Rather we average the inequality (\ref{independence inequality}) over $\Phi$ from some subset of $\cB$ and obtain a less restrictive but more tractable bound, which constitutes \\
%
%
{\bf Theorem 1 (The necessary condition for the system ISI).}
Let the hamiltonian $\hH$ have a nondegenerate energy spectrum.
Let $\cF$ be some (possibly small) subset of a restricted subspace $\cB_R\subset\cB.$ Assume that the equilibrium state $\meandensS$ of the system is independent from the initial state of the system with the accuracy $\varepsilon$ for all initial states of the bath which belong to $\cF.$ Then
\be\label{Theorem 1}
\sup_{\psi\in\cS} \left|\left| ~ \la\meandensS\ra_{\cB_R}[\psi]-\la\meandensS\ra_{\cS\otimes\cB_R} \right|\right|<\varepsilon' 
\ee
with
\be\label{varepsilon'}
\varepsilon'\leq \varepsilon +2 \sqrt{\frac{d_{\cS}}{d_R}}+ \frac{2}{\sqrt[3]{d_R}}+ \frac8p e^{-c \sqrt[3]{d_R}},~~~c=1/(18\pi^3).
\ee
Here $p$ is the measure of $\cF$ (with respect to the uniform normalized measure on $\cB_R$).
In other words, for a random $\Phi\in\cB_R$
$$p=\Pr\left\{ \Phi\in\cF \right\}.$$
Note, however, that the dimensionality of $\cF\subset\cB_R$ should be equal to the dimensionality of $\cB_R,$ otherwise $p=0$ and $\varepsilon'=\infty.$

The {\it proof} of Theorem 1, which is largely based on Theorem 0, may be found in the Appendix.

According to (\ref{varepsilon'}), the subset $\cF$ may have an exponentially small measure $p$, and still  $\varepsilon'$ would be small enough to make the bound (\ref{Theorem 1}) restrictive. Indeed, $\varepsilon'$ is small as long as $\varepsilon$ is small and $d_R$ is sufficiently large to ensure that $p\gg e^{-c \sqrt[3]{d_R}}.$ Thus Theorem 1 states that if the system ISI property holds for at least exponentially small number of the bath initial conditions $\Phi,$ then the restrictive bound (\ref{Theorem 1}) is valid. Usually it is natural to demand that the the system ISI property holds for those initial states of environment which have well-defined energy. In this case the set $\cF$ may be constructed from those $\Phi$ which provide a small dispersion to $\hHB.$

The averages $\la\meandensS\ra_{\cB_R}[\psi]$ and $\la\meandensS\ra_{\cS\otimes\cB_R}$ take more explicit
form in the specific case when $\cB_R=\cB.$ \\
{\bf Lemma.}
\be\label{totally averaged}
\la\meandensS\ra_{\cS\otimes\cB}=(d_{\cS})^{-1}\mathbb{1}^{\cS}.
\ee
If further the hamiltonian $\hH$ has a nondegenerate energy spectrum, then
\be\label{B-averaged}
\la\meandensS\ra_{\cB}[\psi]=
\frac{1}{d_{\cB}} \sum_{n=1}^{d}   \la\psi|\densS_n|\psi\ra \densS_n.
\ee
With this Lemma in hand one may reformulate Theorem 1 to obtain the following.\\
%
%
{\bf Theorem 1$'$.}
Let the hamiltonian $\hH$ have a nondegenerate energy spectrum.
Assume that the equilibrium state $\meandensS$ of the system is independent from the initial state of the system with the accuracy $\varepsilon$ for all initial states of the bath $\Phi$ from some (possibly small) subset $\cF\subset\cB.$ Then
\be\label{Theorem 1'}
\sup_{\psi\in\cS} \left|\left|~\frac{1}{d_{\cB}}\sum_{n=1}^{d}   \la\psi|\densS_n|\psi\ra \densS_n- \frac{\mathbb{1}^{\cS}}{d_{\cS}} \right|\right|<\varepsilon',
\ee
where $\varepsilon'$ is bounded according to (\ref{varepsilon'}) with $d_R=d_{\cB}.$\\
{\it Proof.} Theorem 1$'$ follows directly from Theorem 1 and Lemma. Therefore it is sufficient to prove the Lemma.
As far as $\meandensS[\psi\Phi]$ is a quadratic form with respect both to $\psi$ and to $\Phi,$ the averaging over $\cS$ and $\cB$ with a uniform measure is equivalent to the averaging over arbitrary orthonormal bases in $\cS$ and $\cB$ correspondingly:
\be\label{averaging and tracing}
\la\meandensS[\psi\Phi]\ra_{\cS}=d_{\cS}^{-1}\sum_{j=1}^{d_{\cS}} \meandensS[\psi_j\Phi],
~~~\la\meandensS[\psi\Phi]\ra_{\cB}=d_{\cB}^{-1}\sum_{l=1}^{d_{\cB}} \meandensS[\psi\Phi_l].
\ee
Applying this rule to the decomposition (\ref{meandensS expanded}) and taking into account that
\be
d_\cB^{-1} \sum_{l=1}^{d_\cB} |\la\Psi_n|\psi\Phi_l\ra|^2 = d_\cB^{-1} \la\psi|\densS_n|\psi\ra
\ee
one obtains the equalities (\ref{totally averaged}) and (\ref{B-averaged}).
~$\boxempty$


Although Theorem 1 is stronger than Theorem 1$'$, the latter may be easier applied for the analysis of the  specific models. For this reason we concentrate on Theorem 1$'$ in what follows.

In fact Theorem 1$'$ states that if the system ISI property holds, then the majority of $\densS_n\equiv\tr_{\cB}|\Psi_n\ra\la \Psi_n|$ should be approximately proportional to the unit matrix,
\be\label{densS average}
\densS_n \simeq d_{\cS}^{-1} {\mathbb{1}^{\cS}}.
\ee
This requirement is natural. Indeed, according to \cite{Popescu2006} almost all vectors $\Psi$ from $\cH$ yield  $\tr_{\cB}|\Psi\ra\la \Psi|\simeq d_{\cS}^{-1} {\mathbb{1}^{\cS}}.$ More precisely, for a random vector $\Psi\in\cH$
\be
\Pr\left\{ \left|\left|~\tr_{\cB}|\Psi\ra\la \Psi|-d_{\cS}^{-1} {\mathbb{1}^{\cS}}~\right|\right|> \sqrt{\frac{d_{\cS}}{d_{\cB}}}+\epsilon\right\}\leq 2 e^{-c d_{\cB} \epsilon^2}.
\ee
Therefore, for a generic Hamiltonian $\hH$ one expects the bound (\ref{Theorem 1'}) to hold with a  fairly small $\varepsilon.$

To get more insight in the statement of Theorem 1$'$ let us consider a situation in which the system ISI property is known to hold. Namely, consider the weak interaction case
and assume that thermalization occurs at the level of {\it individual eigenstates} \cite{Deutsch1991}\cite{Srednicki1994}, which means that $\densS_n=Z^{-1}\exp(-\beta_n \hHS)$ for (almost) all $n.$\footnote{The inverse temperature $\beta_n$ for individual eigenstates $\Psi_n$ of the composite system is defined in a usual way, $\beta_n=\left.\frac{d \ln r}{dE}\right|_{E=E_n},$  where the state density function $r(E)$ is reasonably smoothed.
} In fact general considerations and numerical studies suggest that this {\it eigenstate thermalization hypothesis} holds generically, see e.g. \cite{Rigol2008}. In this case all four thermalization properties are valid. In particular, according to eq.(\ref{meandensS expanded}) the equilibrium state of the system is of the Boltzmann-Gibbs canonical form and does not depend on the initial state of the system, provided the initial state of the composite system has a small energy dispersion.\footnote{
In ref. \cite{Gogolin} (Theorem 2.8.1) one may find a quantitative bound on the time-averaged distance between two states of a system corresponding to two different initial states. This bound is restrictive whenever the  eigenstate thermalization hypothesis is valid.
}
Let us make sure that our necessary condition of the system ISI holds in this case.
Note that as far as the dimensionality of $\cH$ is finite, negative temperatures are allowed as well as positive (see \cite{Ramsey1956} for the discussion of statistical physics with negative temperatures). Normally in such situation the inverse temperature is close to zero  for the vast majority of states. This is especially the case when the bath is composed of many weakly interacting subsystems with identical spectrum, as may be shown with the use of the central limit theorem. Thus the major contribution to the average over $\cB$ comes from the states with high temperature, $\beta_n \simeq 0.$  As a results, eq.(\ref{densS average}) is satisfied for the majority of $n$ and the statement of Theorem 1$'$ holds with some small $\varepsilon.$

Here we would like to make the following remark. Although we assume that the dimensionality of the Hilbert space of the bath is finite throughout the present paper, it seems plausible that all our results, in particular, Theorems 1, 1$'$  (and also Theorem 2, see below) may be generalized to the case when $d_{\cB}=\infty.$ Indeed, $d_{\cB}$ does not enter Theorem 1 at all, while it enters Theorem 1$'$ only through the average $(d_{\cB})^{-1} \sum\limits_{n=1}^{d_{\cS}d_{\cB}}   \la\psi|\densS_n|\psi\ra \densS_n,$ which presumably remains well-defined when  $d_{\cB}\rightarrow\infty.$ In this case Theorem 1$'$ in fact provides a necessary condition for the system ISI in the {\it hight-temperature regime}, when it is natural to expect that all eigenstates of $\meandensS$ are equiprobable independently of the initial state of the system.

Although our necessary condition for the system ISI is expected to hold for {\it generic} Hamiltonians $\cH,$ as is clear from the above discussion, it does not hold for some {\it specific} Hamiltonians. This is exemplified in the next section.
A simple model is discussed there when this condition turns out to be restrictive enough to prove the {\it absence} of the system ISI, although the decoherence and the equilibration occur and the bath ISI is present. 

Before we turn to the specific example let us reformulate our general results in the extreme case when the system $\cS$ is represented by a single spin $1/2.$
In this case the equality (\ref{B-averaged}) may be rewritten as
\be
\la\overline\pp\ra_\cB= d^{-1} \sum_{n=1}^{d}  \pp_n (\pp_0,\pp_n),
\ee
while the inequality (\ref{Theorem 1'}) -- as
\be\label{Theorem 1' spin one half}
\sup_{\pp_0} \left|\la\overline\pp\ra_\cB\right|=\sup_{\pp_0} \left|  d^{-1} \sum_{n=1}^{d} \pp_n (\pp_0,\pp_n)\right|<\varepsilon'.
\ee
%
It turns out that one may get rid of supremum in (\ref{Theorem 1' spin one half}) and obtain
the following.\\
\newline
{\bf Theorem $2$.} Consider the hamiltonian $\hH$ with the nondegenerate energy spectrum. Assume that the equilibrium state of the spin $\cS$ is independent from the initial state of the spin $\pp_0$ with the accuracy $\varepsilon$ for all initial states of the bath $\Phi$ from some (possibly small) subset $\cF\subset\cB.$ Then
\begin{description}
\item[(i)]
\be\label{Theorem 2 i}
  \left(\frac{1}{d^2}\sum_{n=1}^{d}\sum_{m=1}^{d} (\pp_n,\pp_m)^2\right)^{1/2}<\sqrt{3}\varepsilon'
\ee
\item[(ii)]
\be\label{Theorem 2 ii}
 \frac{1}{d} \sum_{n=1}^{d} \pp_n^2<3\varepsilon'
\ee
\end{description}
with $\varepsilon'$ bounded according to (\ref{varepsilon'}) with $d_R=d_{\cB}.$ 

The {\it proof} of Theorem 2 may be found in the Appendix.

The second bound in Theorem 2 is weaker but more tractable than the first one. It shows that the purities of eigenstates, $\tr (\densS_n)^2=(1+\pp_n^2)/2,$ should be on average very close to its minimal value $1/2.$ This requirement also enters the sufficient condition for the system ISI, cf. eq.(\ref{minimal purity}).

\section{\label{sect specific model}Specific model}
In this section we concentrate on a specific class of exactly solvable (to some extent) models in which the system is represented by the spin $1/2$ and the above derived necessary condition for the system ISI is not fulfilled. We consider the Hamiltonian
\be\label{model}
\hH = \frac{\omega}{2}\sigma_z+\frac12\sum_{\alpha=x,y,z}\sigma_\alpha \hV_\alpha +\hHB,
\ee
where $\sigma_\alpha$ acts in $\cS,$  $\hV_\alpha$ acts in $\cB,$ at least one of $\hV_{x,y}$ is nontrivial (i.e. not zero and not proportional to the unit operator), all  $\hV_\alpha$ commute with each other,
\be
[\hV_\alpha,\hV_\beta]=0,
\ee
and the interaction Hamiltonian commutes with the bath self-Hamiltonian,
\be
[\hV_\alpha,\hHB]=0 ~~~~~\forall \alpha.
\ee
Note, however, that the interaction Hamiltonian does not commute with the system self-Hamiltonian: $[\hHS,\hHSB]=(\ii/2)(\sigma_y \hV_x-\sigma_x \hV_y)\neq 0.$ This means in particular that the system energy is not a conserved quantity, and one may expect some sort of thermalization of the system.

Let $\Phi_l,~ l=1,...,d_{\cB}$ be the common eigenvectors of $\hV_\alpha,~\alpha=x,y,z,$ and $\hHB:$
\be
\hV_\alpha\Phi_l=v_{l \alpha }\Phi_l,~~~\hHB\Phi_l=E_l^{\cB}\Phi_l.
\ee
The eigenvectors and eigenvalues of the total Hamiltonian
read
\be\label{eigenproblem solution}
\Psi_{l\pm}=\psi_{l\pm} \Phi_l,~~~E_{l\pm}=E_l^{\cB}\pm\frac12\sqrt{(\omega+v_{lz})^2+v_{lx}^2+v_{ly}^2},
\ee
where $\psi_{l\pm}$ are two eigenvectors of the $l-$dependent matrix
$({\omega}\sigma_z+{\bm v_l} \ssigma).$ We assume that the total Hamiltonian has nondegenerate energy gaps (and, consequently, nondegenerate spectrum), which is clearly a generic case.

A specific version of the model under consideration  (with $\hV_y=\hV_z=0$ and the bath composed of noninteracting spins) was introduced  in \cite{Cucchietti2005} in the context of decoherence studies.
It was shown in \cite{Cucchietti2005} that the decoherence occurs effectively in the sense that the spin which is initially in a pure state rapidly becomes entangled with the bath in the course of the evolution.

Equilibration and the bath ISI are also present in the model for almost all initial states of the bath, which follows from the general results of \cite{Linden2009}. Namely, let us fix the initial state of the system $\psi$ and choose some initial state of the bath $\Phi$ from a large bath subspace $\cB_R\subset \cB.$ As was proven in \cite{Linden2009} the time averaged distance between $\densS(t)$ and $\meandensS$ is small,
\be
\overline{||\densS(t)-\meandensS||}\leq 2 \frac{d_{\cS}}{\sqrt{d_R}} = \frac{4}{\sqrt{d_R}}
\ee
for almost all $\Phi\in\cB_R.$ The exceptional $\Phi,$ which violate the above bound, form a set of exponentially small measure. This proves the equilibration property. The bath ISI property may be proven with the use of Theorem 0, see section \ref{sect sufficient condition}.

Thus our model is nontrivial enough to decohere and equilibrate effectively and to  have equilibrium states which are almost independent from the bath initial states. However, it does not match the necessary condition for the system initial state independence imposed by Theorem 2. This is essentially because the eigenstates of the composite system  are factorized, see eq.(\ref{eigenproblem solution}), which results in $|\pp_n|=1$ for any $n.$ 
This ensures that if  $\varepsilon'<\frac13$  then the restriction (\ref{Theorem 2 ii}) imposed by Theorem 2 can not be satisfied. In other words, for the overwhelming majority of the initial states of the bath the equilibrium state $\meandensS$ of the system can not be independent from the initial state of the system $\psi$ with the accuracy better than $1/3.$

Two other examples when the system ISI property is absent were already considered in \cite{Linden2009}. In the first considered case there exist at least one conserved quantity of the system, i.e. a nontrivial operator $A$ which acts in $\cS$ and commutes with the total Hamiltonian $\hH.$ Evidently  equilibrium states are different for different expectation values $\la\psi|A|\psi\ra.$

In the second case the range of energies of the self-Hamiltonian $\hHS$ is greater than the range of energies of the combined interaction-bath Hamiltonian $\hHSB+\hHB:$
$$E^{\cS}_{\rm max}-E^{\cS}_{\rm min}>E^{\cS\cB+\cB}_{\rm max}-E^{\cS\cB+\cB}_{\rm min}.$$
In this case the system can not transfer to (or from) the bath a substantial amount of energy, and the equilibrium state depends on the initial energy of the system (although the energy of the system is not strictly conserved).

Our example differs from the examples provided in \cite{Linden2009}. Indeed, there are no conserved quantity of the system in our model, and the range of energies of the self-Hamiltonian $\hHS$ (which is equal to $\omega$) may be arbitrary small. The distinctive feature of the considered model, which leads to the absence of the system ISI, is total lack of the entanglement of eigenvectors of $\hH.$ Remind that the high degree of entanglement was required to prove the system ISI property in case when $d_{\cS}\gg 1$ \cite{Linden2009} (see the discussion in section \ref{sect sufficient condition}). Now we show that the absence of entanglement leads to the breakdown of the system initial state independence in the opposite case when  $d_{\cS}= 2.$ Thus the entanglement seems to be an indispensable condition for the system ISI.

We emphasize however that the {\it exact} lack of entanglement ($|\pp_n|=1$) is not of key importance in the above considerations which proved the absence of the system ISI property. 
Rather, according to Theorem 2 (ii), the value of the average $\frac{1}{d} \sum_{n=1}^{d} \pp_n^2$ is essential. If it is greater than some $x,$ then the equilibrium state of the system can not be independent from the initial state with the accuracy considerably better than $x/3.$


\section{\label{sect summary}Summary}

To conclude, we have considered the system initial state independence property -- one of the cornerstones of the equilibrium statistical mechanics. We present a {\it necessary condition} for this property to hold (Theorem 1). This condition may be applied in particular in the case when $d_{\cS}$ (the dimensionality  of the Hilbert space of the system which undergoes thermal relaxation)  is small. This case is of special interest, as the {\it sufficient} condition proved previously \cite{Linden2009} does not work for small $d_{\cS}.$

If we demand that the system ISI property holds with a fixed accuracy for the whole range of the bath ``macrostates'' (e.g., for all states of the bath with small energy dispersion), then we get a more explicit form of the necessary condition (Theorem $1'$). The latter indicates that the majority of eigenstates of the total Hamiltonian (which includes self-Hamiltonians of the system and the bath, as well as the interaction term) should be highly entangled.

When the equilibrating system is just a single spin $1/2,$ our necessary condition leads to the transparent bounds on the polarization vectors of the total Hamiltonian eigenstates (Theorem 2).
%
The usefulness of the derived bounds is demonstrated in the specific case. Namely, it is shown that for a specific form of interaction between the spin and the bath the necessary condition is not satisfied and thus the system initial state independence property does not hold. The considered interaction is not completely trivial; in particular it leads to  the decoherence of the spin.
Two other properties which are associated with thermal relaxation -- the equilibration and the bath initial state independence -- also hold in the considered model.

Our results are negative in the sense that they allow only to pinpoint those models which lack the system initial state independence property.  Further work is necessary to obtain more insight in the problem, in particular, to find an efficient {\it sufficient} condition for the system initial state independence in the case when $d_{\cS}$ is small. Also it is desirable to accurately generalize the obtained results to the case when the Hilbert space of the bath is infinitely-dimensional.

\section*{Acknowledgements}
The author is grateful to V.A. Novikov and E.B. Bogomolny for the constructive criticism and useful remarks. The
work was partly supported by the Dynasty Foundation scholarship, RF
President grant NSh-4172.2010.2, RFBR grants 10-02-01398 and 08-02-00494.

\section*{Appendix}
{\it Proof of Theorem 1.} We need to derive the bound (\ref{Theorem 1}) from the following inequality:
\be\label{Theorem 1 proof starting point}
\left|\left|~\meandensS[\psi\Phi]-\la\meandensS\ra_{\cS}[\Phi]~\right|\right|<\varepsilon ~~~\forall\psi\in\cS, ~\forall \Phi\in\cF.
\ee
The latter along with the triangle inequality implies that
\be\label{Theorem 1 proof eq2}
\left|\left|~\la\meandensS\ra_{\cF}[\psi]-\la\la\meandensS\ra_{\cS}\ra_{\cF}~\right|\right|
<\varepsilon~~~\forall\psi\in\cS.
\ee
Now we have to move from averaging over small subset $\cF\subset\cB_R$ to averaging over the whole large $\cB_R.$ From (\ref{Theorem 1 proof eq2}) one gets
\be\label{Theorem 1 proof eq3}
\left|\left|~\la\meandensS\ra_{\cB}[\psi]-\la\la\meandensS\ra_{\cS}\ra_{\cB}~\right|\right|
<\varepsilon+ \left|\left|~\la\meandensS\ra_{\cF}[\psi]-\la\meandensS\ra_{\cB}[\psi]~\right|\right|+ \left|\left|~\la\la\meandensS\ra_{\cS}\ra_{\cF}-\la\la\meandensS\ra_{\cS}\ra_{\cB}~\right|\right|~~~\forall\psi\in\cS.
\ee
Now two last terms in the r.h.s. should be bounded.
First we note that
\be\label{appendix bound 1}
 \left|\left|~\la\meandensS\ra_{\cF}[\psi]-\la\meandensS\ra_{\cB}[\psi]~\right|\right|
  \leq \left\la\left|\left|\meandensS[\psi\Phi]-\la\meandensS\ra_{\cB}[\psi]~\right|\right|\right\ra_{\cF},
\ee
\be\label{appendix bound 2}
 \left|\left|~\la\la\meandensS\ra_{\cS}\ra_{\cF}-\la\la\meandensS\ra_{\cS}\ra_{\cB}~\right|\right|
 \leq \left\la\left\la\left|\left|\meandensS[\psi\Phi]-\la\meandensS\ra_{\cB}[\psi]~\right|\right|\right\ra_{\cF}\right\ra_{\cS}.
\ee
Next we fix some $\psi,$ take some arbitrary $\epsilon>0$ and divide the set $\cF$ in two nonintersecting parts, $\cF_1$ and $\cF_2,$ such as
\be
\cF_2\equiv\left\{\Phi\in\cF: \left|\left|\meandensS[\psi\Phi]-\la\meandensS\ra_{\cB}[\psi]\right|\right|>\sqrt{\frac{d_{\cS}}{d_{R}}}+\epsilon~\right\},
~~~\cF_1\equiv \cF\backslash\cF_2.
\ee
According to Theorem 0 (ii), 
\be
{\rm m(\cF_2)}<2 e^{-c d_R \epsilon^2},
\ee
where $\rm m({\cal A})$ is the measure of the set ${\cal A}$ (remind that we take  $\rm m(\cB_R)=1$).
Evidently, 
\be
\la...\ra_{\cF}=\frac{\rm m(\cF_1)}{\rm m(\cF)}\la...\ra_{\cF_1}+\frac{\rm m(\cF_2)}{\rm m(\cF)}\la...\ra_{\cF_2},
\ee
and one gets
\be\label{appendix bound 3}
\begin{array}{c}
\left\la\left|\left|\meandensS[\psi\Phi]-\la\meandensS\ra_{\cB}[\psi]~\right|\right|\right\ra_{\cF}\leq
\left\la\left|\left|\meandensS[\psi\Phi]-\la\meandensS\ra_{\cB}[\psi]~\right|\right|\right\ra_{\cF_1}+\frac{2}{p} e^{-c d_R \epsilon^2} \left\la\left|\left|\meandensS[\psi\Phi]-\la\meandensS\ra_{\cB}[\psi]~\right|\right|\right\ra_{\cF_2}\\
\leq \sqrt{\frac{d_{\cS}}{d_{R}}}+\epsilon + \frac{4}{p} e^{-c d_R \epsilon^2}
\end{array}
\ee
for any $\psi$ and $\epsilon,$ where the definition  $p\equiv{\rm m(\cF)}$ is taken into account. Now we have to choose the optimal one. Inserting the estimate (\ref{appendix bound 3}) into eqs.(\ref{appendix bound 1}),(\ref{appendix bound 2}) one evaluates the r.h.s. of the inequality (\ref{Theorem 1 proof eq3}) and gets the desired final expression for $\varepsilon'.$
~$\boxempty$
\\
\newline
{\it Proof of Theorem 2.} \\
({\bf i}) To get the bound (\ref{Theorem 2 i}) from eq.(\ref{Theorem 1' spin one half}) one needs to prove that
\be\label{Theorem 2 i proof}
\sup_{\pp_0} K({\rm p}_{0x},{\rm p}_{0y},{\rm p}_{0z}) \geq \frac{1}{3}\sum_{n=1}^{d}\sum_{m=1}^{d} (\pp_n,\pp_m)^2,
\ee
where $K({\rm p}_{0x},{\rm p}_{0y},{\rm p}_{0z})\equiv  \sum_{n=1}^{d}\sum_{m=1}^{d} (\pp_n,\pp_0)(\pp_m,\pp_0)(\pp_n,\pp_m)=d^2 \la\overline\pp\ra_\cB^2$  is a positive semidefinite quadratic form.  One may rotate the basis to make $K$ diagonal:
\be
K({\rm p}_{0x'},{\rm p}_{0y'},{\rm p}_{0z'})=\lambda_{x'}{\rm p}_{0x'}^2+\lambda_{y'}{\rm p}_{0y'}^2+\lambda_{z'}{\rm p}_{0z'}^2, ~~~0\leq\lambda_{x'}\leq\lambda_{y'}\leq\lambda_{z'}.
\ee
The maximal value of $K$ on the unit sphere is $\lambda_{z'}\geq(\lambda_{x'}+\lambda_{y'}+\lambda_{z'})/3=\tr K/3=\frac{1}{3} \sum_{n=1}^{d}\sum_{m=1}^{d} (\pp_n,\pp_m)^2,$ which is exactly the bound (\ref{Theorem 2 i proof}).\\
\newline
({\bf ii}) To derive the bound (\ref{Theorem 2 ii}) from the bound (\ref{Theorem 2 i}) one needs to prove that 
\be\label{Theorem 2 ii proof}
\sum_{n=1}^{d}\sum_{m=1}^{d} (\pp_n,\pp_m)^2 \geq  \frac{1}{3} \left(\sum_{n=1}^{d} \pp_n^2\right)^2.
\ee
Let us consider the l.h.s. of the above inequality as a function of $3d$ variables $\pp_1,...\pp_d$ and find its minimum subject to $d$ constraints of the form $\pp_n^2=a_n,~n=1,...,d,$ where $0\leq a_n\leq 1$ are some fixed numbers. We introducing $d$ Lagrange multipliers $\eta_n$ to get the Lagrange function $L(\pp_1,...,\pp_d,\eta_1,...,\eta_d)=\sum_{n=1}^{d}\sum_{m=1}^{d} (\pp_n,\pp_m)^2+\sum_{n=1}^{d}\eta_n(\pp_n^2-a_n).$ Differentiation of the latter over ${\rm p}_{m\alpha}$ gives $3d$ equations which (along with the $d$ constraints) define the critical points:
\be\label{Lagrange equations}
2\sum_{n=1}^{d} (\pp_n,\pp_m){\rm p}_{n\alpha}=\eta_m {\rm p}_{m\alpha}, ~~~ m=1,...,d,~\alpha=x,y,z.
\ee
Assume that we already know the set of vectors $\pp_n$ which minimize the l.h.s. of eq.(\ref{Theorem 2 ii proof}) subject to the imposed constraints. This set of vectors should obey equations (\ref{Lagrange equations}), which may be rewritten as
\be\label{Theorem 2 ii proof eigenproblem}
\sum_{\beta=x,y,z}M_{\alpha\beta} {\rm p}_{m \beta}=\eta_m {\rm p}_{m\alpha}
\ee
for every $m$ and $\alpha.$ Here $||M_{\alpha\beta}|| \equiv ||2\sum_{n=1}^{d} {\rm p}_{n\alpha} {\rm p}_{n\beta}||$ is a $3\times3$ real symmetric matrix. Note that it does not depend on $m,$ which is of key importance for the present proof. It has three orthonormal eigenvectors $\bm \xi_1,\bm \xi_2,\bm \xi_3.$ According to (\ref{Theorem 2 ii proof eigenproblem}) every nonzero $\pp_n$ is collinear to one of this eigenvectors and, consequently, orthogonal to two other eigenvectors. In other words, in the set of vectors $\pp_n$ which minimize the l.h.s. of eq.(\ref{Theorem 2 ii proof}) subject to the constraints each two vectors are either collinear, or orthogonal. Without loss of generality we assume that $\pp_1,...,\pp_{d_1}$ are collinear with  $\bm \xi_1,$ $\pp_{d_1+1},...,\pp_{d_2}$ are collinear with  $\bm \xi_2,$ and $\pp_{d_2+1},...,\pp_d$ are collinear with  $\bm \xi_3.$
Then 
\be
\sum_{n=1}^{d}\sum_{m=1}^{d} (\pp_n,\pp_m)^2=\left( \sum_{n=1}^{d_1} a_n \right)^2+\left( \sum_{n=d_1+1}^{d_2} a_n \right)^2+\left( \sum_{n=d_2+1}^{d} a_n \right)^2\geq \frac13 \left( \sum_{n=1}^{d} a_n \right)^2,
\ee
which proves the inequality (\ref{Theorem 2 ii proof}).
~$\boxempty$

\end{document}